\documentclass[runningheads]{llncs}
\usepackage[utf8]{inputenc}
\usepackage{graphicx}
\usepackage{amsmath}
\usepackage{amssymb}
\usepackage{float}
\usepackage{nicefrac}  
\usepackage{color}
\usepackage{cite}
\usepackage{appendix}

\newcommand\numberthis{\addtocounter{equation}{1}\tag{\theequation}}

\usepackage{graphicx}

\begin{document}

\title{Capsule Networks as Generative Models}
\author{Alex B. Kiefer$^*$ \inst{1, 2} \and
Beren Millidge$^*$ \inst{1,3} \and
Alexander Tschantz$^*$ \inst{1} \and\\
Christopher L Buckley\inst{1,4}}
\authorrunning{Kiefer, Millidge, Tschantz, et al}
\institute{VERSES Research Lab \and
Monash University \and
MRC Brain Network Dynamics Unit, University of Oxford \and
Sussex AI Group, Department of Informatics, University of Sussex }
\maketitle

\begin{abstract}
Capsule networks are a neural network architecture specialized for visual scene recognition. Features and pose information are extracted from a scene and then dynamically routed through a hierarchy of vector-valued nodes called ‘capsules’ to create an implicit scene graph, with the ultimate aim of learning vision directly as inverse graphics.  Despite these intuitions, however, capsule networks are not formulated as explicit probabilistic generative models; moreover, the routing algorithms typically used are ad-hoc and primarily motivated by algorithmic intuition. In this paper, we derive an alternative capsule routing algorithm utilizing iterative inference under sparsity constraints. We then introduce an explicit probabilistic generative model for capsule networks based on the self-attention operation in transformer networks and show how it is related to a variant of predictive coding networks using Von-Mises-Fisher (VMF) circular Gaussian distributions.
\end{abstract}

\section{Introduction}
Capsule networks are a neural network architecture designed to accurately capture and represent part-whole hierarchies, particularly in natural images \cite{hinton2011transforming,sabour2017dynamic,hinton2018matrix}, and have been shown to outperform comparable CNNs at visual object classification, adversarial robustness, and ability to segment highly overlapping patterns  \cite{sabour2017dynamic,hinton2018matrix}. A capsule network comprises layers of `capsules' where each capsule represents both the identity and existence of a visual feature as well as its current `pose' (position, orientation, etc.) relative to a canonical baseline.

This approach is heavily inspired by the concept of a scene graph in computer graphics, which represents the objects in a scene in precisely such a hierarchical tree structure where lower-level objects are related to the higher-level nodes by their pose. The capsule network aims to flexibly parameterize such a scene graph as its generative model and then perform visual object recognition by inverting this generative model \cite{hinton2011transforming,hinton2021represent} to infer 3D scene structure from 2D appearances.

It is argued that the factoring of scene representations into transformation-equivariant capsule activity vectors (i.e. vectors that change linearly with translation, rotation, etc.) and invariant pose transformation matrices is more flexible and efficient than the representation used in convolutional neural networks, where activities in higher layers are merely invariant to changes in viewpoint. 
In addition to arguably providing a better scene representation, capsule networks can use agreement between higher-level poses and their predictions based on lower-level poses to solve the binding problem of matching both the `what' and the `where' of an object or feature together.

Capsule networks are in part motivated by the idea that `parse-trees' of the object hierarchy of a scene must be constructed at run-time, since they can be different for different images. Crucially, it is assumed that this dynamically constructed parse-tree must be sparse and almost singly connected - each low-level capsule or feature can be matched to only one high-level parent. This is because in natural scenes it is sensible to assume that each feature only belongs to one object at a time -- for instance, it is unlikely that one eye will belong to two faces simultaneously. In \cite{sabour2017dynamic}, it is proposed to dynamically construct these parse-trees by an algorithm called `routing by agreement' whereby low-level capsules are assigned to the high-level capsule whose pose matrix most closely matches their pose matrix under certain transformations.

While capsule networks appear to be a highly efficient architecture, invented using deep insights into the nature of visual scenes, there are, nevertheless, many elements of the construction that appear relatively ad-hoc. There is no construction of an explicit probabilistic generative model of the network. Moreover, it is unclear why the routing algorithm works and how it is related to other frameworks in machine learning. Indeed, some research \cite{Paik2019CapsuleNN, Rawlinson2018SparseUC} suggests that typical routing algorithms do not perform well which suggests that the goals of routing are better attained in some other way.

In this paper we propose a probabilistic interpretation of capsules networks in terms of Gaussian mixture models and VMF (circular Gaussian) distributions, which applies the self-attention mechanism used in modern transformer networks \cite{vaswani2017attention,gregor2015draw}. We argue that fundamentally, the purpose of the original routing-by-agreement algorithm of \cite{sabour2017dynamic} is to approximate posterior inference under a generative model with the particular sparsity structure discussed above. We first demonstrate in experiments that we can achieve routing-like behaviour using sparse iterative inference, and show in addition that even in the original implementation of dynamic routing in capsules \cite{sabour2017dynamic}, sparsity of the top-level capsules is enforced via the margin loss function alone when iterative routing is turned off. This loss function can be interpreted as implementing a low-entropy prior on digit classes. We then write down a principled top-down generative model for capsules networks that provides a plausible description of the model that routing attempts to approximately invert. Overall, our results aim to provide a clear and principled route toward understanding capsule networks, and interpreting the idea of routing as fundamentally performing sparse iterative inference to construct sparse hierarchical program trees at runtime -- a method that can be implemented in many distinct ways.
\section{Capsule Networks}
A capsule network comprises a hierarchical set of layers each of which consists of a large number of parallel capsules. In practice, several non-capsule layers such as convolutional layers are often used to provide input data preprocessing. We do not consider non-capsule layers in this analysis. 

Each capsule $j$ in a layer receives an input vector $\mathbf{s}_j$ consisting of a weighted combination of the outputs of the capsules $i$ in the layer below, multiplied by their respective affine transformation matrices $\mathbf{T}_{i,j}$, which define the invariant relationships between the poses represented by $i$ and $j$. The input from capsule $i$ is denoted $\hat{u}_{j|i} = \mathbf{T}_{i,j}v_i$, where $v_i$ is the output activity of capsule $i$ after its input has been passed through a `squash' nonlinearity defined as $f(\mathbf{x}) = \frac{||\mathbf{x}||^2}{1 + ||\mathbf{x}||^2}\cdot \frac{\mathbf{x}}{||\mathbf{x}||}$. The higher-level capsule then weights the contributions from its low-level input capsules by weighting coefficients $c_{i,j}$ which are determined by iterative routing. To obtain the output of the capsule, all its inputs are weighted and summed and then the output is fed through the nonlinear activation function $f$. The forward pass of a capsule layer can thus be written as,
\begin{align}
    v_{{(l)_j}} = f(\sum_i c_{i,j} \mathbf{T}_{i,j} v_{{(l-1)}_i})
    \label{capsule_forward_pass}
\end{align}

The core algorithm in the capsule network is the routing-by-agreement algorithm which iteratively sets the agreement coefficients $c_{i,j}$:

\begin{equation}
\begin{aligned}
    b^k_{i,j} &= b^{k-1}_{i,j} + (\mathbf{T}_{i,j}\mathbf{v}_{{(l-1)}_i})^T{\mathbf{v}_{{(l)}_j}^{k-1}} \\[10pt]
    \mathbf{c}^k_{i} &= \sigma(\mathbf{b}^{k-1}_i)
\label{routing_iteration}
\end{aligned}
\end{equation}
where $k$ is the iteration index of the routing algorithm, $\sigma(\mathbf{x})$ is the softmax function such that $\sigma(\mathbf{x})_i = \frac{\exp(x_i)}{\sum_j \exp(x_j)}$, and $\mathbf{b}^{k}_i$ are the logit inputs to the softmax at iteration $k$, which act as log priors on the relevance of lower-level capsule $i$'s output to all the higher-level capsules. These are initialized to $0$ so all capsules are initially weighted equally. 

The routing algorithm weights the lower-level capsule's contribution to determining the activities at the next layer by the dot-product similarity between the input from the low-level capsule and the higher-level capsule's output at the previous iteration. Intuitively, this procedure will match the pose of the higher-level capsule and the `projected pose' $\mathbf{\hat{u}}_{j|i}=\mathbf{T}_{i,j} \mathbf{v}_{{(l-1)}_i}$ from the lower-level capsule, so that each lower-level capsule will predominantly send its activity to the higher-level capsule whose pose best fits its prediction. In addition to matching parts to wholes, this procedure should also ensure that only higher-level capsules that receive sufficiently accurate pose-congruent `votes' from the capsules below are activated, leading to the desired sparsity structure in the inferred scene representation.

\section{Sparse Capsule PCN}
Intuitively, the goal of routing is to match the poses of higher- and lower-level capsules and thus to construct a potential parse tree for a scene in terms of relations between higher-level `objects' and lower-level 'features'. Crucially, this parse tree must be highly sparse such that, ideally, each lower-level feature is bound to only a single high-level object. To represent uncertainty, some assignment of probability to other high-level capsules may be allowed, but only to a few alternatives. 

We argue that all of this is naturally accommodated if we interpret routing as implementing \emph{sparse iterative inference}, where the sparsity constraints derive from an implicit underlying generative model. This is because the fundamental goal of routing is to obtain a `posterior' over the capsule activations throughout the network given the input as `data'. Unlike standard neural networks, this posterior is not only over the classification label at the output but over the `parse tree' comprising activations at the intermediate layers. Taking inspiration from Predictive Coding Networks (PCNs) \cite{buckley2017free,bogacz2017tutorial,friston2005theory,millidge2021predictive}, we can imagine the parse tree posterior as being inferred in a principled way through iterative variational inference \cite{wainwright2008graphical,beal_variational_2003} applied to the activities at each layer during a single stimulus presentation. 

The idea of using variational inference to perform capsule routing is also explored and shown to be very effective in \cite{ribeiro2020capsule}. Most closely related to our aims here, \cite{DBLP:journals/corr/abs-2004-03553} propose a full generative model and variational inference procedure for capsules networks, focusing instead on the E-M routing version of capsules \cite{hinton2018matrix} in which existence is explicitly represented using a distinct random variable. They show that performing iterative inference to further optimize solutions at test time leads to improved digit reconstructions for rotated MNIST digits.  \cite{DBLP:journals/corr/abs-2103-06676} also proposes a generative model that aims to capture the intuitions behind capsule networks, and likewise derives a variational inference scheme for inverting this model.

There are various ways to achieve sparsity. \cite{Rawlinson2018SparseUC} investigated unsupervised versions of capsules networks, and found that while routing in the CapsNet architecture did not produce the intended effects (i.e. sparse activations at each capsule layer and feature equivariance) without supervision at the output layer, these properties could be restored by adding a sparsity constraint adapted from k-sparse autoencoders \cite{Makhzani2014kSparseA}. A `lifetime sparsity constraint' that forces all capsules to be active a small fraction of the time was also found to be necessary to discourage solutions in which a small number of capsules are used to reconstruct the input and the rest are ignored (which interferes with the ability to learn the desired equivariances). We experiment with a simpler form of sparsity in combination with iterative PC inference, using an L1 penalty, which is known to encourage sparsity, as an additional regularizing term added to the free energy. In Appendix B, we demonstrate this effect on a toy two-layer network where sparse iterative inference routs all inputs through a specific intermediate layer, thus constructing a single preferred parse-tree.

\subsection{Experiments}

\begin{figure}[H]
    \vspace{-0.5cm}
    \centering
    \includegraphics[width=\linewidth]{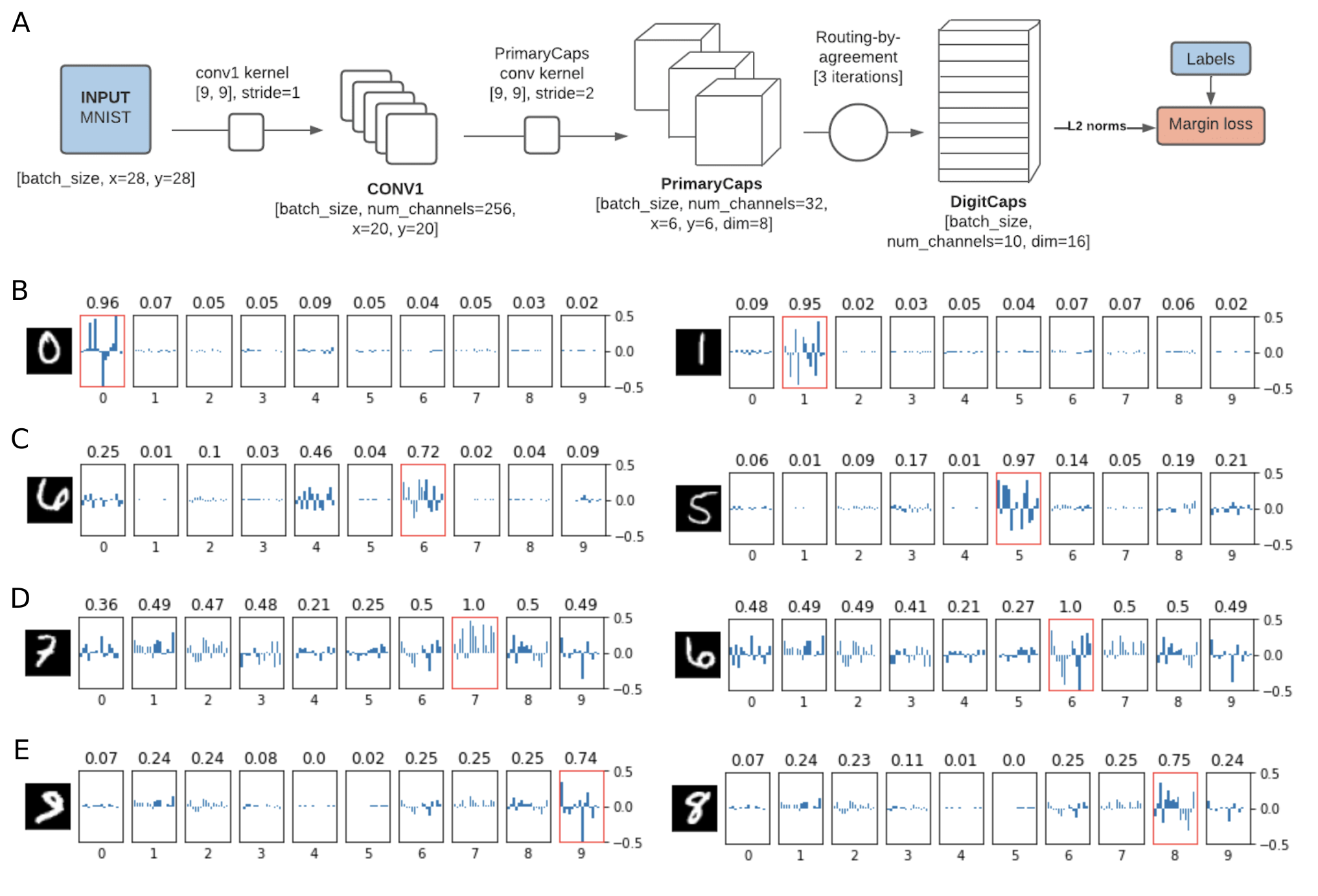}
    \caption{A: CapsNet architecture. Input (left) is passed through Conv1, yielding 256 feature maps which provide the input to the first (convolutional) capsules layer, PrimaryCaps. This yields 1152 (32 dimensions x 6 * 6 output) 8-dimensional capsules, which are each connected to each of the 10 16-dimensional DigitCaps via their own transformation matrices. The L2 norms of the digit capsules are then compared with a one-hot encoding of the target. The auxiliary reconstruction network is not pictured. Rows B-E: Samples of DigitCaps activity vectors for test set examples under varying conditions. Red borders indicate correct digits and the number above each box is the corresponding vector norm. B: DigitCaps activities using a network trained with three iterations of dynamic routing. Sparsity is clearly enforced at the capsule level, though it is more extreme in some cases than others. C: Two random activity vectors from an otherwise identical network trained without routing. Note that the ambiguous image on the left predictably leads to less decisive capsule outputs (note also that this occurrence was not unique to the no-routing condition). D: Capsule network trained without routing, with 500 iterations of iterative inference performed in place of routing at inference time. E: Same as (D) but with an L1 regularization term on the capsule activities (i.e. $\Sigma_j{\|\mathbf{v}_j\|}$) added to the standard predictive coding (squared prediction error) loss function.}
    \label{capsnet_fig}
    \vspace{-0.5cm}
\end{figure}

To test our interpretation of iterative routing-by-agreement as inference under sparsity constraints, we investigated the role of routing in the canonical `CapsNet' capsules network proposed in \cite{sabour2017dynamic}. This network, diagrammed in Fig. \ref{capsnet_fig}A, consists of a preliminary conventional convolutional layer, followed by a convolutional capsules layer, and a final layer whose ten capsules are meant to represent the presence of the digits 0-9 in input images. Three iterations of the routing-by-agreement algorithm are used between the two capsules layers.

In the experiments reported in \cite{sabour2017dynamic}, the network is trained to classify MNIST digits via backpropagation, using a separate `margin loss' function for each digit:
\begin{align}
    \mathcal{L}_k = \mathcal{T}_k\max{(0,m^+ -\lVert\mathbf{v}_k \rVert})^2 + \lambda(1 -\mathcal{T}_k)\max{(0,\lVert \mathbf{v}_k \rVert - m^-})^2
    \label{margin_loss}
\end{align}
Here, $\mathcal{L}_k$ is the margin loss for digit $k$ ($0$ through $9$), $\mathcal{T}_k$ is a Boolean indicating the presence of that digit in the input image, $\lVert \mathbf{v}_k \rVert$ is the L2 norm of capsule output $\mathbf{v}_k$, and $m^+$ and $m^-$ are thresholds used to encourage these vector norms to be close to 1 or 0, in the digit-present or digit-absent conditions, respectively. $\lambda$ is an additional term used to down-weight the contribution of negative examples to early learning. In the full CapsNet architecture, this loss is combined with a downweighted reconstruction regularization term from an auxiliary reconstruction network used to encourage capsule activities to capture relevant features in the input.

In our experiments, we first trained a standard CapsNet on MNIST for 13 epochs, using the same hyperparameters and data preprocessing as in \cite{sabour2017dynamic}. It is worth noting that, as a baseline for comparison, CapsNet does indeed produce sparse higher-level capsule activations (i.e. sparsity in the vector of L2 norms of the activity vectors of each of the 10 capsules in the DigitCaps layer). However, in Fig. \ref{capsnet_fig}C we show that training the same network architecture with $0$ routing iterations (simply setting the higher-level capsule activities to the squashed sum of the unweighted predictions from lower layers) produces sparsity as well, suggesting that the transformation matrices learn to encode this sparsity structure based on the margin loss function alone. In addition to exhibiting similar higher-level capsule activities, the two networks also performed comparably (with $> 99\%$ accuracy on the test set) after training, though the no-routing network was initially slower to converge (see Appendix A).

To test whether the intuitions explored in the toy example in Appendix B would play out in a larger-scale architecture, we also tried using iterative inference in place of dynamic routing. In these experiments, we began with a forward pass through a trained CapsNet, clamped the target (label) nodes, and ran iterative inference for 500 iterations with a learning rate of $0.01$, either with the standard squared-error predictive coding objective or with the standard PC loss plus L1 regularization applied to the final output vector of the CapsNet (see Appendix C). We found that, as in the toy experiment, standard iterative inference with an L1 penalty (in this case applied per-capsule) produced sparse outputs, while without the L1 penalty activity was more evenly distributed over the capsules, though the vector norms for the `correct' capsules were still longest.

Overall, our findings on CapsNet are consistent with results reported in \cite{Rawlinson2018SparseUC}, which suggest that the sparsity seen at the output layer of CapsNet is attributable to its supervised learning objective alone and does not occur without this objective. Further confirming these results, we performed experiments on a modification of CapsNet with an intermediate capsule layer between PrimaryCaps and DigitCaps, and did not observe sparsity in the intermediate-layer activities despite comparable performance. Despite our largely negative findings, these experiments support our broader view that the main point of routing is to induce sparsity in the capsule outputs, and that this objective can be achieved by various means, including iterative inference in a predictive coding network. In the following section we propose an explicit generative model for capsules that produces the right kind of sparsity in a principled way.
\section{A Generative Model for Capsules}
To be fully consistent with the goal of learning vision as inverse computer graphics, a capsules network should be formulated as a top-down model of how 2D appearances are generated from a hierarchy of object and part representations, whose inversion recovers a sensible parse-tree. We now develop an explicit probabilistic generative model for the capsule network that achieves this which, interestingly, involves the self-attention mechanism used in transformer networks.

\subsection{Attention and the Part-Whole Hierarchy}

In recent years it has become increasingly clear that neural attention \cite{vaswani2017attention,gregor2015draw} provides the basis for a more expressive class of artificial neural network that incorporates interactions between activity vectors on short timescales. As noted in \cite{sabour2017dynamic}, while conventional neural networks compute their feedforward pass by taking the dot products of weight vectors with activity vectors, neural attention relies on the the dot product between two activity vectors, thus producing representations that take short-term context into account. In particular, attention allows for the blending of vectors via a weighted sum, where the weights depend on dot-product similarities between input and output vectors. The core computation in neural attention can be written as,
\begin{align}
    \mathbf{Z} = \sigma(\mathbf{Q}\mathbf{K}^T)\mathbf{V}
    \label{self_attention_formula}
\end{align}
with $\mathbf{Z}$ being the output of the attention block,  $\mathbf{K},\mathbf{Q},\mathbf{V}$ being the `Key', `Query', and `Value' matrices, and $\sigma$ the softmax function, as above. Intuitively, the attention operation can be thought of as first computing `similarity scores' between the query and key matrices and then normalizing them with the softmax. The similarity scores are then multiplied by the value matrix to get the output. In the transformer architecture \cite{vaswani2017attention,bahdanau2014neural}, these matrices are typically produced from a given input representation (e.g. a word embedding) via a learned linear transformation.

There is a tempting analogy between the capsule layer update and neural attention, since the output capsule activities are determined as a blend of functions of the inputs, using weights determined by applying the softmax function to dot-product similarity scores. A key difference, which seems to ruin the analogy, is that in routing-by-agreement, each of the weights that determine the output mixture comes from a distinct softmax, over the outputs of one lower-level capsule. Simply swapping out a neural attention module for the routing algorithm gives the wrong result however, since this enforces a `single-child constraint' where in the limit each higher-level object is connected to at most one lower-level part in the parse-tree.

It turns out however that the attention mechanism is precisely what is needed to naturally construct a \emph{top-down} generative model of parse trees within a capsules architecture. Firstly, we note that we can aggregate the small transformation matrices  $\mathbf{T}_{i,j}$ connecting input capsule $i$ to output capsule $j$ into a large tensor structure $\mathbf{W}=\begin{bmatrix}\mathbf{T}_{1,1} & \dots & \mathbf{T}_{m,1} \\ \vdots & \ddots & \vdots \\ \mathbf{T}_{1,n} & \dots & \mathbf{T}_{m,n}\end{bmatrix}$, for $m$ lower-level and $n$ higher-level capsules. Similarly, the $N$ individual $d$-dimensional capsule vectors in a layer can be stacked to form an $N \times d$ matrix with vector-valued entries, $\mathbf{V}_{(l)} = [{\mathbf{v}_{(l)}}_1, {\mathbf{v}_{(l)}}_2 \dots {\mathbf{v}_{(l)}}_N]^T$, and the routing coefficients $c_{ij}$ collected into a matrix $\mathbf{C}$ with the same shape as $\mathbf{W}$. We can then write the forward pass through a vector of capsules in a way that is analogous to a forward pass through a large ANN:
\begin{align}
    \label{mlp_description}
    \mathbf{V}_{(l)} &= f\Big[(\mathbf{C}\odot \mathbf{W})\mathbf{V}_{(l-1)}\Big]
\end{align}
Here, $\odot$ denotes element-wise multiplication and the nonlinearity $f$ is also applied element-wise. The expression $\mathbf{W}\mathbf{V}_{(l-1)}$ should be read as a higher-level matrix-matrix multiplication in which matrix-vector multiplication is performed in place of scalar multiplication per element, i.e. ${\mathbf{V}_{(l)}}_j = {\sum_{i=1}^{m}}{\mathbf{C}_{ji}\mathbf{W}_{ji}\mathbf{V}_{{(l-1)}_{i}}}$. This term implements the sum of predictions ${\sum_{i=1}^m}\mathbf{\hat{u}}_{j|i}$ from lower-level capsules for the pose of each higher-level capsule $j$, where each transformation matrix is first scaled by the appropriate entry in the routing coefficient matrix $\mathbf{C}$.

We have argued that what the forward pass in equation \ref{mlp_description} aims to implement is in effect posterior inference under a top-down generative model. To write down such a model, we first define $\mathbf{\tilde{W}}$ as the transpose of the original weight tensor $\mathbf{W}$, i.e. an $m \times n$ collection of transformations from $n$ higher-level capsules to $m$ lower-level capsules. Since each row of $\mathbf{\tilde{W}}$ collects the transformations from all higher-level capsules to the intrinsic coordinate frame of one lower-level capsule ${\mathbf{v}_{(l-1)}}_i$, we can then define a matrix $\mathbf{\hat{U}}_{(l-1)} = \begin{bmatrix}
({\mathbf{\tilde{W}}}_1\odot\mathbf{V}_{(l)})^T\\
\vdots \\
({\mathbf{\tilde{W}}}_m\odot\mathbf{V}_{(l)})^T\\
\end{bmatrix}$ that contains all the predictions for the lower-level capsules, where matrix-vector multiplication is applied element-wise to the submatrices.

Setting $\mathbf{V} = \mathbf{K} = \mathbf{\hat{U}}_{{(l-1)}_i}$ and $\mathbf{Q} = \mathbf{V}_{{(l-1)}_i}$, we can then frame the top-down generation of one lower-level capsule vector within a capsules network as an instance of neural attention, where $\mathbf{V}^k_{(l)}$ is the matrix of capsule activities at layer $l$ and iteration $k$, as above:
\begin{align}
    {\mathbf{V}^k}_{{(l-1)}_i} = \sigma({\mathbf{V}^{k-1}_{{(l-1)}_i}}{\mathbf{\hat{U}}_{{(l-1)}_i}}^T)\mathbf{\hat{U}}_{{(l-1)}_i}
    \label{capsules_attention_formula}
\end{align}
These updates are independent for each lower-level capsule but can clearly be vectorized by adding an extra leading dimension of size $m$ to each matrix, so that an entire attention update is carried out per row.

The above can be viewed as an inverted version of routing-by-agreement in which the higher-level capsules cast `votes' for the states of the lower-level capsules, weighted by the terms in the softmax which play a role similar to routing coefficients. There is also a relation to associative memory models \cite{krotov2020large,ramsauer2020hopfield,millidge2022universal} since we can think of the capsule network as associating the previous lower-level output (query) to `memories' consisting of the predictions from the layer above, and using the resulting similarity scores to update the outputs as a blend of the predictions weighted by their accuracy. 

Crucially, when used in this way for several recurrent iterations, neural attention encourages each row of the output matrix to be dominated by whichever input it is most similar to. Since the output in this case is the lower level in a part-whole hierarchy (where rows correspond to capsule vectors), this precisely enforces the single-parent constraint that routing-by-agreement aspires to. 

In the routing-by-agreement algorithm, the routing logits $b_{ij}$ are initially set to $0$ (or to their empirical prior value, if trained along with the weights) and then accumulate the dot-product similarities during each iteration of routing. It is clear that the application of attention alone without the accumulation of log evidence over iterations should produce a routing-like effect, since there is a positive feedback loop between the similarities and the softmax weights. If one wanted to emulate routing-by-agreement more closely, the above could be supplemented with an additional recurrent state formed by the similarity scores of the previous iteration.

While the single-parent constraint alone is not sufficient to ensure that only lower-level capsules that are a good fit with some active higher-level capsule are activated, it is reasonable to expect that when no constant relationship between a higher- and lower-level entity exists in the data, training would encourage the weights of the corresponding transformation matrix to be close to $0$, on pain of inappropriately activating a lower-level capsule, which would lead to an increase in the loss function (e.g. squared prediction error).

\subsection{Probabilistic Generative Model}
We now formulate the generative model sketched above explicitly in probabilistic terms, which therefore also doubles as a generative model of transformer attention (see Appendix D).

As remarked above, capsule networks can be seen as combining something like a standard MLP forward pass with additional vector-level operations. In particular, for a given lower-level capsule $i$, the attention mechanism can be interpreted as mixing the MLPs defined by each capsule pair $(i, j)$ with the weights given by the attention softmax. If we interpret each MLP update in terms of a Gaussian distribution, as in PCNs (that is, where the uncertainty about the hidden state $\mathbf{V}_{(l-1)}$ is Gaussian around a mean given by the sum of weighted `predictions'), we arrive at a mixture of Gaussians distribution over each lower-level capsule pose, whose mixing weights are determined by scaled dot-product similarity scores.

These similarity scores must be interpreted differently from the MLP-like part of the model, and we argue that these are correctly parametrized via a von Mises-Fisher (VMF) distribution with a mean of the pose input. The VMF implements the dot-product similarity score required by the routing coefficient in a probabilistic way, and can be thought of as parametrizing a distribution over vector angles between the output and input poses, which is appropriate since the purpose of the routing coefficients is to reinforce predictions that match a higher-level pose (where degree of match can be captured in terms of the angle between the pose and prediction vectors). Importantly, since it parametrizes an \emph{angle} the distribution is circular since angles `wrap-around'. The VMF distribution is a Gaussian defined on the unit hypersphere and so correctly represents this circular property.

Given the above, we can express the update in equation \ref{capsules_attention_formula} above as a probabilistic generative model as follows:
\begin{align*}
    p({\mathbf{V}_{{(L)}_i}|\mathbf{\hat{U}}_{{(L)}_i}}) &=
    {\sum_{j}\Big[{\pi_{{(i)}_j}}{\mathcal{N}(\mathbf{V}_{{(L)}_i}; \mathbf{\hat{U}}_{{(L)}_{ij}}}, \sigma_{ij})}\Big]
    \\[10pt]
    \mathbf{\pi}_{(i)} &= Cat(n, \mathbf{p_{(i)}}) \\[10pt]
    \mathbf{p}_{(i)} &= \sigma{
        \begin{bmatrix}
            VMF(\mathbf{V}_{{(L)}_i}; \mathbf{\hat{U}}_{{(L)}_{i1}}, \kappa_{i1}) & \dots & VMF(\mathbf{V}_{{(L)}_i}; \mathbf{\hat{U}}_{{(L)}_{in}}, \kappa_{in}) 
        \end{bmatrix}
    }
    \numberthis
    \label{single_capsule_probs}
\end{align*}
where $\pi_{(i)}$ are the mixing weights, $\sigma_{ij}$ is the standard deviation of the Gaussian distribution over capsule $i$ conditioned on higher-level capsule $j$, and $\kappa_{ij}$ is the `concentration parameter' of the corresponding VMF distribution, which determines how tightly probability mass is concentrated on the direction given by the mean $\mathbf{\hat{U}}_{{(L)}_{ij}}$.

The generative model then defines the conditional probability of an entire capsule layer given the predictions as the product of these per-capsule mixture distributions, mirroring the conditional independence of neurons within a layer in conventional MLPs:
\begin{align}
    p(\mathbf{V_{(L)}}|\mathbf{\hat{U}}_{(L)}) = \prod_i p(\mathbf{V}_{{(L)}_i}|\mathbf{\hat{U}}_{{(L)}_i})
\end{align}
It would be simpler to use the attention softmax itself directly to determine the mixing weights of each GMM, but using the probabilities returned by individual VMF distributions instead affords a fully probabilistic model of this part of the generative process, where not only the vector angle match but also the variance can be taken into account for each capsule pair $i, j$. It remains for future work to write down a process for inverting this generative model using variational inference.

\section{Conclusion}
In this paper, we have aimed to provide a principled mathematical interpretation of capsule networks and link many of its properties and algorithms to other better known fields. Specifically, we have provided a probabilistic generative model of the capsule network in terms of Gaussian and VMF distributions which provides a principled mathematical interpretation of its core computations. Secondly, we have shown how the ad-hoc routing-by-agreement algorithm described in \cite{sabour2017dynamic} is related to self-attention. Moreover, we have demonstrated both in a toy illustrative example and through large-scale simulation of capsule networks how the desiderata of the routing algorithm can be achieved through a general process of sparse iterative inference.

\section*{Acknowledgements}
Alex Kiefer and Alexander Tschantz are supported by VERSES Research.
Beren Millidge is supported by the BBSRC grant BB/S006338/1 and by VERSES Research. 
CLB is supported by BBRSC grant number BB/P022197/1 and by Joint Research with the National Institutes of Natural Sciences (NINS), Japan, program No. 0111200.

\section*{Code Availability}

Code for the capsules network is adapted from \\
\text{https://github.com/adambielski/CapsNet-pytorch} \\and can be found at:
\text{https://github.com/exilefaker/capsnet-experiments}.
Code reproducing the toy model experiments and figure in Appendix B can be found at: \text{https://github.com/BerenMillidge/Sparse\_Routing}.

\bibliographystyle{splncs04}
\bibliography{cites}

\begin{thebibliography}{10}
\providecommand{\url}[1]{\texttt{#1}}
\providecommand{\urlprefix}{URL }
\providecommand{\doi}[1]{https://doi.org/#1}

\bibitem{bahdanau2014neural}
Bahdanau, D., Cho, K., Bengio, Y.: Neural machine translation by jointly
  learning to align and translate. arXiv preprint arXiv:1409.0473  (2014)

\bibitem{beal_variational_2003}
Beal, M.J.: Variational algorithms for approximate {Bayesian} inference. Tech.
  rep. (2003)

\bibitem{bogacz2017tutorial}
Bogacz, R.: A tutorial on the free-energy framework for modelling perception
  and learning. Journal of mathematical psychology  \textbf{76},  198--211
  (2017)

\bibitem{bricken2021attention}
Bricken, T., Pehlevan, C.: Attention approximates sparse distributed memory.
  arXiv preprint arXiv:2111.05498  (2021)

\bibitem{brown2020language}
Brown, T.B., Mann, B., Ryder, N., Subbiah, M., Kaplan, J., Dhariwal, P.,
  Neelakantan, A., Shyam, P., Sastry, G., Askell, A., et~al.: Language models
  are few-shot learners. arXiv preprint arXiv:2005.14165  (2020)

\bibitem{buckley2017free}
Buckley, C.L., Kim, C.S., McGregor, S., Seth, A.K.: The free energy principle
  for action and perception: A mathematical review. Journal of Mathematical
  Psychology  \textbf{81},  55--79 (2017)

\bibitem{buzsaki2014log}
Buzs{\'a}ki, G., Mizuseki, K.: The log-dynamic brain: how skewed distributions
  affect network operations. Nature Reviews Neuroscience  \textbf{15}(4),
  264--278 (2014)

\bibitem{chen2021decision}
Chen, L., Lu, K., Rajeswaran, A., Lee, K., Grover, A., Laskin, M., Abbeel, P.,
  Srinivas, A., Mordatch, I.: Decision transformer: Reinforcement learning via
  sequence modeling. Advances in neural information processing systems
  \textbf{34},  15084--15097 (2021)

\bibitem{de2011spatiotemporal}
De~Zeeuw, C.I., Hoebeek, F.E., Bosman, L.W., Schonewille, M., Witter, L.,
  Koekkoek, S.K.: Spatiotemporal firing patterns in the cerebellum. Nature
  Reviews Neuroscience  \textbf{12}(6),  327--344 (2011)

\bibitem{demircigil2017model}
Demircigil, M., Heusel, J., L{\"o}we, M., Upgang, S., Vermet, F.: On a model of
  associative memory with huge storage capacity. Journal of Statistical Physics
   \textbf{168}(2),  288--299 (2017)

\bibitem{dosovitskiy2020image}
Dosovitskiy, A., Beyer, L., Kolesnikov, A., Weissenborn, D., Zhai, X.,
  Unterthiner, T., Dehghani, M., Minderer, M., Heigold, G., Gelly, S., et~al.:
  An image is worth 16x16 words: Transformers for image recognition at scale.
  arXiv preprint arXiv:2010.11929  (2020)

\bibitem{friston2005theory}
Friston, K.: A theory of cortical responses. Philosophical transactions of the
  Royal Society B: Biological sciences  \textbf{360}(1456),  815--836 (2005)

\bibitem{graham2006sparse}
Graham, D.J., Field, D.J.: Sparse coding in the neocortex. Evolution of nervous
  systems  \textbf{3},  181--187 (2006)

\bibitem{greff2016highway}
Greff, K., Srivastava, R.K., Schmidhuber, J.: Highway and residual networks
  learn unrolled iterative estimation. arXiv preprint arXiv:1612.07771  (2016)

\bibitem{gregor2015draw}
Gregor, K., Danihelka, I., Graves, A., Rezende, D., Wierstra, D.: Draw: A
  recurrent neural network for image generation. In: International conference
  on machine learning. pp. 1462--1471. PMLR (2015)

\bibitem{hinton2021represent}
Hinton, G.: How to represent part-whole hierarchies in a neural network. arXiv
  preprint arXiv:2102.12627  (2021)

\bibitem{hinton2011transforming}
Hinton, G.E., Krizhevsky, A., Wang, S.D.: Transforming auto-encoders. In:
  International conference on artificial neural networks. pp. 44--51. Springer
  (2011)

\bibitem{hinton2018matrix}
Hinton, G.E., Sabour, S., Frosst, N.: Matrix capsules with em routing. In:
  International conference on learning representations (2018)

\bibitem{jastrzkebski2017residual}
Jastrzbski, S., Arpit, D., Ballas, N., Verma, V., Che, T., Bengio, Y.: Residual
  connections encourage iterative inference. arXiv preprint arXiv:1710.04773
  (2017)

\bibitem{kanerva1988sparse}
Kanerva, P.: Sparse Distributed Memory. MIT Press (1988)

\bibitem{krotov2020large}
Krotov, D., Hopfield, J.: Large associative memory problem in neurobiology and
  machine learning. arXiv preprint arXiv:2008.06996  (2020)

\bibitem{krotov2016dense}
Krotov, D., Hopfield, J.J.: Dense associative memory for pattern recognition.
  Advances in Neural Information Processing Systems  \textbf{29},  1172--1180
  (2016)

\bibitem{lamme2000distinct}
Lamme, V.A., Roelfsema, P.R.: The distinct modes of vision offered by
  feedforward and recurrent processing. Trends in neurosciences
  \textbf{23}(11),  571--579 (2000)

\bibitem{Makhzani2014kSparseA}
Makhzani, A., Frey, B.J.: k-sparse autoencoders. CoRR  \textbf{abs/1312.5663}
  (2014)

\bibitem{melloni2012interaction}
Melloni, L., van Leeuwen, S., Alink, A., M{\"u}ller, N.G.: Interaction between
  bottom-up saliency and top-down control: how saliency maps are created in the
  human brain. Cerebral cortex  \textbf{22}(12),  2943--2952 (2012)

\bibitem{millidge2022universal}
Millidge, B., Salvatori, T., Song, Y., Lukasiewicz, T., Bogacz, R.: Universal
  hopfield networks: A general framework for single-shot associative memory
  models. arXiv preprint arXiv:2202.04557  (2022)

\bibitem{millidge2021predictive}
Millidge, B., Seth, A., Buckley, C.L.: Predictive coding: a theoretical and
  experimental review. arXiv preprint arXiv:2107.12979  (2021)

\bibitem{DBLP:journals/corr/abs-2103-06676}
Naz{\'{a}}bal, A., Williams, C.K.I.: Inference for generative capsule models.
  CoRR  \textbf{abs/2103.06676} (2021), \url{https://arxiv.org/abs/2103.06676}

\bibitem{olshausen1996emergence}
Olshausen, B.A., Field, D.J.: Emergence of simple-cell receptive field
  properties by learning a sparse code for natural images. Nature
  \textbf{381}(6583),  607--609 (1996)

\bibitem{olshausen2004sparse}
Olshausen, B.A., Field, D.J.: Sparse coding of sensory inputs. Current opinion
  in neurobiology  \textbf{14}(4),  481--487 (2004)

\bibitem{Paik2019CapsuleNN}
Paik, I., Kwak, T., Kim, I.: Capsule networks need an improved routing
  algorithm. ArXiv  \textbf{abs/1907.13327} (2019)

\bibitem{parmar2018image}
Parmar, N., Vaswani, A., Uszkoreit, J., Kaiser, L., Shazeer, N., Ku, A., Tran,
  D.: Image transformer. In: International conference on machine learning. pp.
  4055--4064. PMLR (2018)

\bibitem{pearl1988probabilistic}
Pearl, J.: Probabilistic reasoning in intelligent systems: networks of
  plausible inference. Morgan kaufmann (1988)

\bibitem{radford2019language}
Radford, A., Wu, J., Child, R., Luan, D., Amodei, D., Sutskever, I., et~al.:
  Language models are unsupervised multitask learners. OpenAI Blog
  \textbf{1}(8), ~9 (2019)

\bibitem{ramsauer2020hopfield}
Ramsauer, H., Sch{\"a}fl, B., Lehner, J., Seidl, P., Widrich, M., Adler, T.,
  Gruber, L., Holzleitner, M., Pavlovi{\'c}, M., Sandve, G.K., et~al.: Hopfield
  networks is all you need. arXiv preprint arXiv:2008.02217  (2020)

\bibitem{Rawlinson2018SparseUC}
Rawlinson, D., Ahmed, A., Kowadlo, G.: Sparse unsupervised capsules generalize
  better. ArXiv  \textbf{abs/1804.06094} (2018)

\bibitem{reed2022generalist}
Reed, S., Zolna, K., Parisotto, E., Colmenarejo, S.G., Novikov, A.,
  Barth-Maron, G., Gimenez, M., Sulsky, Y., Kay, J., Springenberg, J.T.,
  et~al.: A generalist agent. arXiv preprint arXiv:2205.06175  (2022)

\bibitem{ribeiro2020capsule}
Ribeiro, F.D.S., Leontidis, G., Kollias, S.D.: Capsule routing via variational
  bayes. In: AAAI. pp. 3749--3756 (2020)

\bibitem{sabour2017dynamic}
Sabour, S., Frosst, N., Hinton, G.E.: Dynamic routing between capsules.
  Advances in neural information processing systems  \textbf{30} (2017)

\bibitem{schweighofer2001unsupervised}
Schweighofer, N., Doya, K., Lay, F.: Unsupervised learning of granule cell
  sparse codes enhances cerebellar adaptive control. Neuroscience
  \textbf{103}(1),  35--50 (2001)

\bibitem{shepherd2018handbook}
Shepherd, G.M., Grillner, S.: Handbook of brain microcircuits. Oxford
  University Press (2018)

\bibitem{DBLP:journals/corr/abs-2004-03553}
Smith, L., Schut, L., Gal, Y., van~der Wilk, M.: Capsule networks - {A}
  probabilistic perspective. CoRR  \textbf{abs/2004.03553} (2020),
  \url{https://arxiv.org/abs/2004.03553}

\bibitem{sterling2015principles}
Sterling, P., Laughlin, S.: Principles of neural design. MIT press (2015)

\bibitem{theeuwes2010top}
Theeuwes, J.: Top--down and bottom--up control of visual selection. Acta
  psychologica  \textbf{135}(2),  77--99 (2010)

\bibitem{tschantz2022hybrid}
Tschantz, A., Millidge, B., Seth, A.K., Buckley, C.L.: Hybrid predictive
  coding: Inferring, fast and slow. arXiv preprint arXiv:2204.02169  (2022)

\bibitem{vanrullen2007power}
VanRullen, R.: The power of the feed-forward sweep. Advances in Cognitive
  Psychology  \textbf{3}(1-2), ~167 (2007)

\bibitem{vaswani2017attention}
Vaswani, A., Shazeer, N., Parmar, N., Uszkoreit, J., Jones, L., Gomez, A.N.,
  Kaiser, {\L}., Polosukhin, I.: Attention is all you need. In: Advances in
  Neural Information Processing Systems. pp. 5998--6008 (2017)

\bibitem{wainwright2008graphical}
Wainwright, M.J., Jordan, M.I., et~al.: Graphical models, exponential families,
  and variational inference. Foundations and Trends{\textregistered} in Machine
  Learning  \textbf{1}(1--2),  1--305 (2008)

\bibitem{weidner2009sources}
Weidner, R., Krummenacher, J., Reimann, B., M{\"u}ller, H.J., Fink, G.R.:
  Sources of top--down control in visual search. Journal of Cognitive
  Neuroscience  \textbf{21}(11),  2100--2113 (2009)

\bibitem{willmore2011sparse}
Willmore, B.D., Mazer, J.A., Gallant, J.L.: Sparse coding in striate and
  extrastriate visual cortex. Journal of neurophysiology  \textbf{105}(6),
  2907--2919 (2011)

\bibitem{zheng2022online}
Zheng, Q., Zhang, A., Grover, A.: Online decision transformer. arXiv preprint
  arXiv:2202.05607  (2022)

\end{thebibliography}

\section*{Appendix A: Convergence of CapsNet with and without routing}

The following plots show the loss per epoch (plotted on a log scale for visibility) during training of a CapsNet architecture with 3 and 0 rounds of dynamic routing-by-agreement and without an auxiliary reconstruction net. The figure shows that after initially slower learning, CapsNet without routing converged to nearly the same test set loss as with routing.

\begin{figure}[h]
    \centering
    \includegraphics[width=\linewidth]{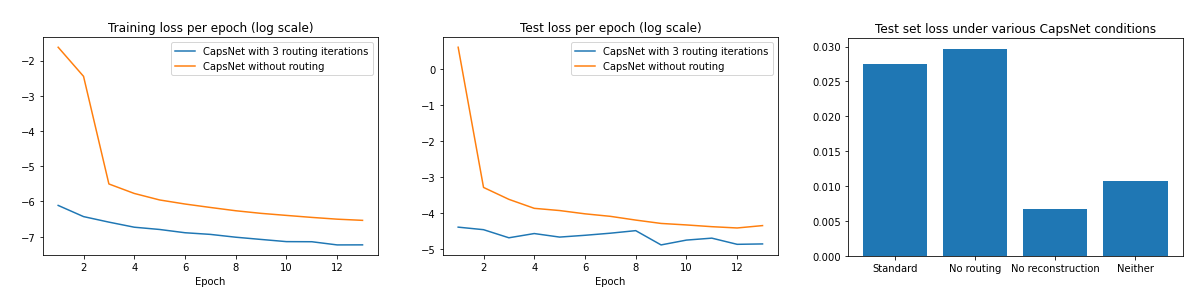}
    \caption{Left: Training set loss during training of CapsNet under standard routing (3 iterations) and no-routing conditions. Middle: Test set loss. Right: Comparison of final losses across standard, no-routing, no-reconstruction, and no-routing no-reconstruction conditions. Note that the loss functions differ between the reconstruction and no-reconstruction conditions.}
    \label{loss_plots}
\end{figure}

Interestingly, although classification performance was very similar across these networks, the test set accuracy for the four conditions (standard, no routing, routing without reconstruction loss, and neither routing nor reconstruction loss) were 99.23\%, 99.34\%, 99.32\%, and 99.29\% respectively. In this case at least, dynamic routing appears not to have led to improved accuracy, although it does lead to slightly lower values of the loss function both when using the full loss (margin + reconstruction) and when using margin loss alone.

This is consistent with the findings in \cite{Paik2019CapsuleNN} that iterative routing does not greatly improve the performance of capsules networks and can even lead to worse performance, though it is also consistent with Sabour et al \cite{sabour2017dynamic}, who report a roughly $0.14\%$ performance improvement using routing against a no-routing baseline.

\section*{Appendix B: Toy Model of Routing as Sparse Iterative Inference}

\begin{figure}
    \centering
    \includegraphics[width=\linewidth]{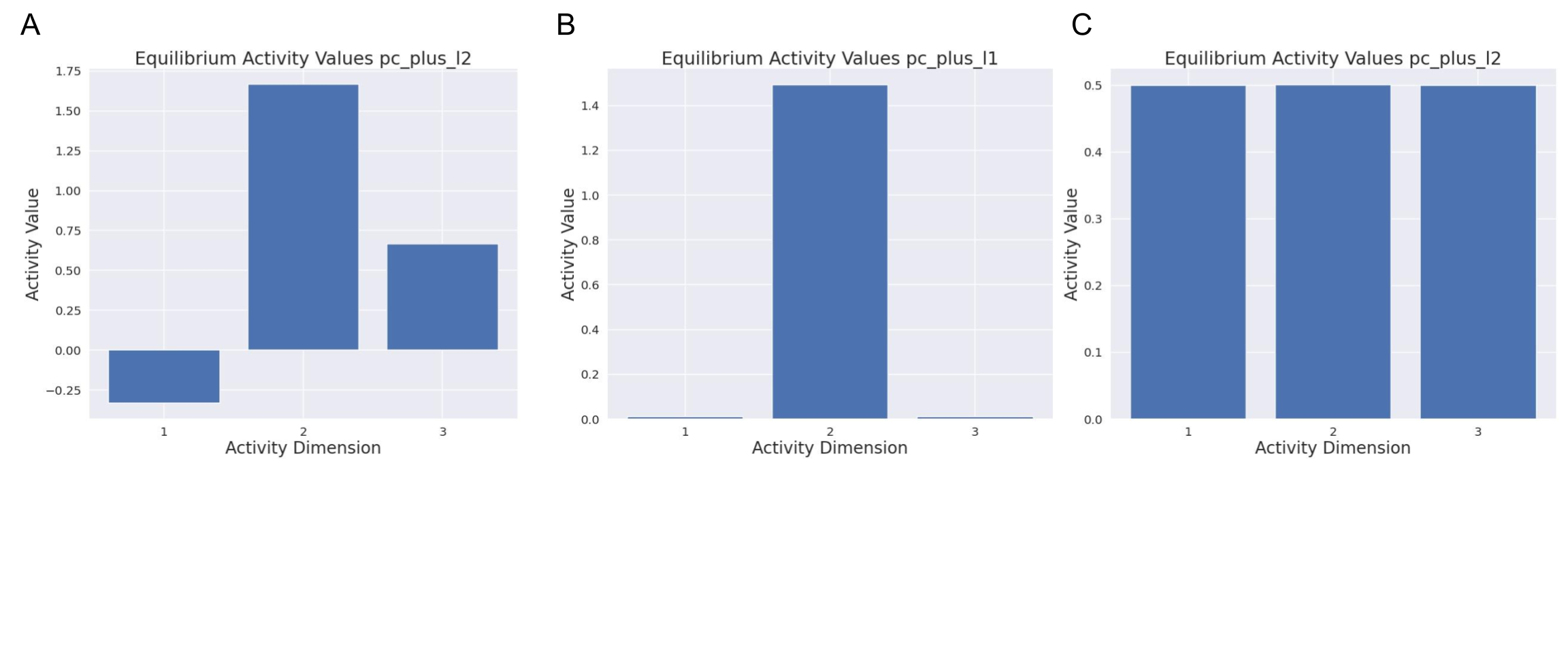}
    \vspace{-2cm}
    \caption{A: Default behaviour of the outcome of iterative inference purely minimizing squared prediction errors. Probability mass is distributed between all three potential `objects' in the hidden layer. B: Outcome of sparse iterative inference using an L1 penalty term in the objective function. All probability mass is concentrated on a single `best' object so that a singly connected scene parse tree is constructed. C: Outcome of iterative inference with an L2 penalty term in the loss function which encourages probability mass to spread out. All objects have approximately equal probability of being selected.}
    \label{toy_model_fig}
    \vspace{-0.5cm}
\end{figure}

Capsule networks assume that a sparse parse tree representation of a stimulus is preferred and achieve this using the routing algorithm while an equivalent ANN would typically produce dense representations. To gain intuition for why sparse iterative inference may be able to achieve this result as well as capsule routing, we provide a simple illustrative example of how iterative inference with a sparsity penalty can result in routing-like behaviour. We consider a simple three-layer neural network with a single hidden layer and visible input and output layers. We fix both the top and bottom layers to an input or a target respectively. We can then infer the hidden layer activities which can both sufficiently `explain' the output given the input. If we imagine the input layer of the network as representing features and the output as a classification label, then in the hidden layer we wish to uniquely assign the input features all to the best matching `object'. We construct such a network with input size 3, hidden size 3, and output size 1, with input weights set to identity and the output weights set to a matrix of all $1$s. The network is linear although this is not necessary. Following the Gaussian generative model proposed for the capsule network, we implemented this network as a predictive coding network (PCN) and performed iterative inference by updating activities to minimize the variational free energy which can be expressed as a sum of squared prediction errors at each layer \cite{buckley2017free}. In additional to the standard iterative free energy, we also experimented with adding either a sparsity penalty (L1 regularisation) or L2 activity norm regularization to the network. We investigated the extent to which iterative inference with the sparsity penalty can reproduce the desired routing effect with only a single high-level feature being active, and found that it can (see Figure \ref{toy_model_fig}). 

Moreover, this sparsity penalty was necessary in this network in order for inference to exhibit routing-like behaviour. Without any regularization, iterative inference has a tendency to distribute probability mass between various high-level objects. This tendency is exacerbated with L2 regularisation which encourages the inference to spread probability mass as evenly as possible.

Interestingly, a similar intuition is applied in \cite{hinton2018matrix} where routing is explicitly derived as part of an EM algorithm with a clear probabilistic interpretation and where the MDL penalties derived for simply activating a capsule, which do not depend on its degree of activation, can perhaps also be thought of as effectively implementing a similar sparsity penalty which encourages the EM algorithm to assign capsule outputs to a single high-level capsule instead of spreading probability mass between them.

\section*{Appendix C: Iterative inference process for CapsNet}

Our implementation of sparse iterative inference in place of capsule routing is the same in outline as that used for the toy model discussed in Appendix B, applied to the CapsNet architecture. That is, we minimize the sum of squared prediction errors per layer, which is equivalent to the variational free energy \cite{millidge2021predictive}. In this case the prediction error for the output layer is given by the difference between the prediction from the penultimate (PrimaryCaps) layer and the clamped target values. For the sparsity condition, we also add the capsule-level L1 sparsity penalty discussed in the caption of figure \ref{capsnet_fig} at the output layer. Dynamic routing was turned off for this experiment, both at inference time and during training.

\section*{Appendix D: Relationship between Capsule Routing and Attention}

As noted above, our generative model of the capsule network can also describe the self-attention block in transformers, providing a fundamental building block towards building a full transformer generative model. Explicitly writing down such a generative model for the transformer architecture could enable a significantly greater understanding of the core mechanisms underlying the success of transformers at modelling large-scale sequence data as well as potentially suggest various improvements to current architectures.

This relationship to transformer attention is important because transformer attention is well-understood and found to be highly effective in natural language processing tasks \cite{radford2019language,brown2020language} as well as recently in vision \cite{parmar2018image,dosovitskiy2020image} and reinforcement learning \cite{chen2021decision,zheng2022online,reed2022generalist}. Since capsule networks appear highly effective at processing natural scene statistics, this provides yet another example of the convergence of machine learning architectures towards a universal basis of attention mechanisms. 

The basis of attention mechanisms can then be further understood in terms of associative memory architectures based on Modern Hopfield networks \cite{krotov2020large,ramsauer2020hopfield}, as briefly discussed above. It has been found that sparsity of similarity scores is necessary for effective associative memory performance to prevent retrieved memories from interfering with each other \cite{kanerva1988sparse,krotov2016dense,millidge2022universal}. The softmax operation in self-attention can be interpreted as a separation function with the goal of sparsifying the similarity scores by exponentially boosting the highest score above the others. Indeed, it is a general result that the capacity of associative memory models can be increased dramatically by using highly sparsifying separation functions such as high-order polynomials \cite{krotov2016dense,demircigil2017model}, softmaxes \cite{ramsauer2020hopfield} and top-k activation functions \cite{bricken2021attention}.

An interesting aspect of our generative model is the use of VMF distributions to represent the dot-product similarity scores. Intuitively, this arises because the cosine similarity is `circular' in that angles near 360 degrees are very similar to angles near 0. In most transformer and associative memory models, the update rules are derived from Gaussian assumptions which do not handle the wrap-around correctly and hence may be subtly incorrect for angles near the wrap-around point. By deriving update rules directly from our generative model, it is possible to obtain updates which handle this correctly and which may therefore perform better in practice. A second potential improvement relates to the VMF variance parameter $\kappa$. In transformer networks this is typically treated as a constant and set to $\frac{1}{\sqrt{d}}$. In essence, this bakes in the assumption that the variance of the distribution is inversely proportional to the data dimension. Future work could also investigate dynamically learning values of $\kappa$ from data which could also improve performance.

One feature of routing-by-agreement not captured by iterative inference in standard PCNs is the positive feedback loop, in which low prediction error encourages even closer agreement between activities and predictions. This is similar to applying self-attention over time. A key distinction between attention as used in transformers and the routing mechanism in capsule networks is that the latter is iterative and can be applied sequentially for many iterations (although usually only 3-5), unlike in transformers where it is applied only once. Capsule networks therefore could provide ideas for improving transformer models by enabling them to work iteratively and adding the recurrent state that arises from the `bias' term in the routing algorithm.

It has been proposed that highly deep networks with residual connections, a set of architectures that includes transformers, are implicitly approximating iterative inference using depth instead of time \cite{greff2016highway,jastrzkebski2017residual} which is a highly inefficient use of parameters. Instead, it is possible that similar performance may be obtained with substantially smaller models which can explicitly perform iterative inference similar to capsule networks. Some evidence for this conjecture comes from the fact that empirically it appears that large language models such as GPT2 \cite{radford2019language} appear to perform most of their decisions as to their output tokens in their first few layers. These decisions are then simply refined over the remaining layers -- a classic use-case for iterative inference.

The link between capsule routing and sparse iterative inference also has significant resonances in neuroscience. It is known that cortical connectivity and activations are both highly sparse (approximately only 1-5\% neurons active simultaneously) \cite{buzsaki2014log,graham2006sparse,willmore2011sparse} with even higher levels of sparsity existing in other brain regions such as the cerebellum \cite{schweighofer2001unsupervised,de2011spatiotemporal,shepherd2018handbook}. Such a level of sparsity is highly energy efficient \cite{sterling2015principles} and may provide an important inductive bias for the efficient parsing and representation of many input signals which are generated by highly sparse processes -- i.e. dense pixel input is usually only generated by a relatively small set of discrete objects. Secondly, iterative inference is a natural fit for the ubiquitous recurrent projections that exist in cortex \cite{theeuwes2010top,weidner2009sources,melloni2012interaction,lamme2000distinct,vanrullen2007power} and many properties of visual object recognition in the brain can be explained through a hybrid model of a rapid amortized feedforward sweep followed by recurrent iterative inference \cite{tschantz2022hybrid}. These considerations combine to provide a fair bit of evidence towards a routing-like sparse iterative inference algorithm being an integral part of cortical functioning. Moreover, it has been demonstrated many times in the sparse-coding literature that adding sparse regularisation on a variety of reconstruction and classification objectives can result in networks developing receptive fields and representations that resemble those found in the cortex \cite{olshausen1996emergence,olshausen2004sparse,willmore2011sparse}.

Iterative inference is also important for enabling object discrimination and disambiguation in highly cluttered and occluded scenes because it can model the vital `explaining away' \cite{pearl1988probabilistic} effect where inferences about one object can then inform parallel inferences about other objects. This is necessary in the case of occlusion since by identifying the occluder and implicitly subtracting out its visual features, it is often possible to make a much better inference about the occluded object \cite{hinton2011transforming}. It is therefore noteworthy, and suggestive of our hypothesis that routing can really be interpreted as iterative inference, that capsule networks perform much better at parsing such occluded scenes than purely feedforward models such as CNNs.

\end{document}